\documentclass[12pt,epsf]{article}
\usepackage{graphicx,float,subfigure,epsfig,color,rotating,latexsym}
\usepackage[latin1]{inputenc}
\usepackage[T1]{fontenc}
\def\be{\begin{equation}}
\def\ee{\end{equation}} 
\def\bea{\begin{eqnarray}}
\def\eea{\end{eqnarray}} 
\def\ba{\begin{array}} 
\def\ea{\end{array}}

\def\ln{{\rm ln}}

\begin{document}

\begin{center} {\bf \large{Time evolution of $ T_{\mu\nu} $ and the cosmological  
constant problem} }\\

\vspace*{0.8 cm}

Vincenzo Branchina\footnote{Vincenzo.Branchina@ct.infn.it}\label{one}

\vspace*{0.4 cm}

Department of Physics, University of
Catania and \\ INFN, Sezione di Catania, 
Via Santa Sofia 64, I-95123, Catania, Italy 

\vspace*{0.8 cm}

Dario Zappal\`a\footnote{Dario.Zappala@ct.infn.it}\label{two}

\vspace*{0.4 cm}

INFN, Sezione di Catania, Via Santa Sofia 64, I-95123, Catania, Italy  

\vspace*{1.2 cm}

{\LARGE Abstract}\\

\end{center}

\vspace*{0.3cm}

We study the cosmic time evolution of an effective quantum field theory  
energy-momentum tensor $T_{\mu\nu}$ and show that, as a consequence 
of the effective nature of the theory, $T_{\mu\nu}$ 
is such that the vacuum energy decreases with time. 
We find that the zero point energy 
at present time is washed out by the cosmological evolution. 
The implications of this finding for the cosmological 
constant problem are investigated.
\vspace*{0.5cm}


\section{Introduction}

A generic feature of systems with an infinite (very large) number of
degrees of freedom is that fluctuations at arbitrarily close points
are independent. When computing physical quantities, this results in
the appearance of divergent terms. This is the case of quantum
field theories, where such terms are generated as soon as the quantum
fluctuations are taken into account. In particular, the calculation of 
zero point energies leads to divergences whose leading term,
when using a momentum cutoff $\Lambda$, goes as $\Lambda^4$. According 
to standard analysis, these terms contribute to the cosmological constant.

One sometimes takes the point of view that the divergences
have no physical meaning and that the definition of the theory has to  
be completed
by some appropriate renormalization procedure that allows
to remove them. In this perspective, the regularization is just
a mathematical step in the calculation of observable quantities.

From a deeper physical point of view, however, it is more satisfactory
to consider a quantum field theory as an effective theory valid up
to a certain scale $\Lambda$, which takes the meaning of ``scale of new
physics'', and consider a hierarchy of theories each having higher and higher
energy range of validity \cite{wilsrep}. This
hierarchical structure is usually believed to end at the Planck
scale $M_P$ where a different theory, most probably string theory, is
supposed to replace ordinary quantum field theories and should
account for the unification of gravity with the other interactions.

As for the zero point energies, the physical meaning of the divergences
is deeply rooted in the underlying harmonic oscillator structure of
a quantum field theory; this is automatically lost if we cancel out
those terms with the help of a formal procedure such as normal
ordering \cite{dewitt}.

Another important ingredient in the formulation of a relativistic
quantum field theory is the selection of the ground state, which is
done by referring to the Lorentz symmetry. According to 
\cite{cpt},  a Lorentz invariant vacuum  $ |0> $ is characterised 
by the requirement that ~ $\hat P_\mu |0>\,=\,0$, where 
~$\hat P_\mu$~ is the field four-momentum operator.  
As clearly explained in \cite{zeldov2} and \cite{weinb}, 
however, this statement is too restrictive.
This is easily seen if we consider the energy-momentum
tensor of a perfect fluid:
$T_{\mu\nu}= (\rho + p)\,u_\mu u_\nu - \rho g_{\mu\nu}$ (where $u_\mu$  
is the fluid
four-velocity, $p$ the pressure and $\rho$ the energy density). In 
order to have a Lorentz invariant vacuum, all we need is the vacuum
expectation value of the energy-momentum tensor operator 
$\hat T_{\mu\nu}$ to be of the form: 

\be\label{emt}
<0|\hat T_{\mu\nu}|0> \, =  \,-\rho \, g_{\mu\nu}\,.
\ee

\noindent
Eq.(\ref{emt}) contains $\hat P_\mu |0>\, = 0$ as a special case. However, 
it is more general and allows for the presence of vacuum condensates.

On the cosmological side, the importance of the quantum field theoretic
contribution to the energy momentum tensor that appears in
the Einstein equations was firstly recognised in \cite{zeldov}
and \cite{zeldov2}. 
In accordance with the idea that the divergences are unphysical 
and have to be discarded, the divergent terms which do not respect the 
constraint imposed by Eq.(\ref{emt}) were removed with the help of a 
renormalization procedure (more precisely, Pauli-Villars regulators were 
used in \cite{zeldov2}). Such a formal approach is 
thoroughly analysed and criticised in \cite{dewitt}. Still, a popular 
prescription (often used nowadays) for the automatic (yet formal) 
cancellation of these divergences is the dimensional regularization 
scheme. In this respect, see \cite{birrdav} (and also \cite{akhmedov} 
and \cite{sirlin}).

In the present work, we would like to pursue a different point of view.
To begin with, we consider an effective field theory with momentum cut-off 
$\Lambda$, where $\Lambda$ is taken to coincide with the Planck scale
$M_P$, and compute the thermal average $<<\hat T_{\mu\nu}>>$ of the 
energy momentum tensor operator. $<<\hat T_{\mu\nu}>>$ contains 
two additive contributions: 
\be\label{enmomt}
<<\hat T_{\mu\nu}>> \,= T_{\mu\nu}^{\,m} + T_{\mu\nu}^{\,v}\, ,
\ee
where $T_{\mu\nu}^{\,v}$ is the vacuum expectation value 
of $\hat T_{\mu\nu}$ (the superscript $v$ stands 
for ``vacuum''), while $T_{\mu\nu}^{\,m}$ corresponds to the 
equilibrium thermal average of the field excitations 
above the vacuum at temperature $T$ (the superscript $m$ stands 
for ``matter''). For weakly interacting 
fields, $T_{\mu\nu}^{\,m}$ can be regarded  as the thermal average 
of the energy momentum tensor of a gas of non interacting particles. 
At $T=0$, one clearly has $T_{\mu\nu}^{\,m}=0$. 

The vacuum contribution $T_{\mu\nu}^{\,v}$ is of special interest 
to our analysis. In fact, due to the well known form of the Planck 
(or Fermi-Dirac) distribution, $T_{\mu\nu}^{\,m}$ is finite and does not 
contain any reference to the physical cut-off $M_P$. On the contrary, 
$T_{\mu\nu}^{\,v}$ contains terms  proportional to $M_P^4$\,, \,$m^2\,M_P^2$ 
and $m^4\,\,ln\,M_P$, where $m$ is the particle mass (see Eqs.(\ref{rho}) 
and (\ref{press}) below). 

According to our effective field theory point 
of view, 
in the r.h.s.  of the Einstein equation, 
\be\label{einstein}
G_{\mu\nu} - \lambda g_{\mu\nu}= 8\,\pi\, G \,T_{\mu\nu}\, ,
\ee 
we consider for $T_{\mu\nu}$ the full contribution coming 
from Eq.(\ref{enmomt}), i.e. we take 
\be\label{tmunu} 
T_{\mu\nu} \, \equiv \,<<\hat T_{\mu\nu}>> \, =  \,T_{\mu\nu}^{\,m} + 
T_{\mu\nu}^{\,v}\, , 
\ee 

\noindent
without discarding any of the terms that appear in this equation.  
Finally, starting at the Planck time $t=t_{P}$, we follow the cosmic 
evolution of $<<\hat T_{\mu\nu}>>$, in particular of 
$T_{\mu\nu}^{\,v}$, with the help of the corresponding 
Friedman equations. 

Let us call $\rho^v$ the vacuum energy density 
and $p^v$ the vacuum pressure. 
Had we considered a renormalization scheme such as dimensional or 
Pauli-Villars regularization, the coefficient $w$ in the equation of 
state (EOS) $p^v=w\,\rho^v$ (after discarding the divergent terms) 
would have been  $w=-1$\, \cite{zeldov2,birrdav,akhmedov,sirlin}.
Accordingly, $\rho^v$ would not evolve with time and could be properly 
interpreted as the vacuum (or zero-point) energy contribution to the 
cosmological constant. This is the standard view.

Within our effective field theory approach, however, where we keep the 
large but finite terms proportional to $M_P^4$ and $M_P^2$, we get a 
different EOS for $p^v$ and $\rho^v$. As we shall discuss in Sect. 2, 
in this case it turns out that  $w=1/3$. This clearly results in a totally 
different cosmic time evolution for $\rho^v$, which will be analyzed in Sect. 3.
Some considerations on the zero point energy of effective field theories are
presented in Sect. 4 while the conclusions are contained in Sect. 5.

\section{Effective field energy-momentum tensor}

Let us begin by considering a free real single component scalar 
field theory. The energy-momentum operator is:
\be\label{phiemt}
\hat T_{\mu\nu} = \partial_\mu\phi\partial_\nu\phi- g_{\mu\nu}{\cal L} 
=\partial_\mu\phi\partial_\nu\phi
- g_{\mu\nu} \left (\frac12\partial_\mu\phi
\partial^\mu\phi -\frac12m^2\phi^2 \right )\,,\ee
where  ${\cal L}$ is the corresponding Lagrangian density.

After considering the standard Fourier decomposition of $\phi$
in creation and annihilation operators $a_{\vec k}^{\dagger}$ and
$a_{\vec k}$, the energy-momentum tensor $T_{\mu\nu}$ of Eq.(\ref{tmunu})
(i.e. the energy-momentum tensor that appears in the r.h.s. of the 
Einstein equation (\ref{einstein})) is obtained by taking the thermal 
average of (\ref{phiemt}) for a statistical equilibrium distribution
at temperature $T$. The non-diagonal terms vanish, 
while the diagonal ones take the form:

\bea\label{diagemt}
T_{_{0\,0}}&=&<<\hat T_{_{0\,0}}>>\, = \frac{1}{V}
\sum_{\vec k}\sum_n <n|\varrho_{_T}|n>\,n_{\vec k}\,\,
\omega_{\vec k} \,+\, \frac{1}{V} \sum_{\vec k}
\frac{\omega_{\vec k}}{2}\,
\\
 T_{ii}&=&<<\hat T_{ii}>>\, = \frac{1}{V}
\sum_{\vec k}\sum_n <n|\varrho_{_T}|n>\,n_{\vec k} \,\,
\frac{(k^i)^2}{\omega_{\vec k}}\,+\, \frac{1}{V}
\sum_{\vec  k}\frac{(k^i)^2}{2\,\omega_{\vec k}}\, , \label{diagemt1}
\eea
where $<<..>> $ indicates the quantum-statistical average, $|n>$ is a 
compact notation for the generic element of the Fock  space basis, 
$\varrho_{_T}$ is the density operator at temperature $T$,\break 
$n_{\vec k} = <n|a_{\vec k}^{\dagger} a_{\vec k}|n>$\,,\, 
$\omega_{\vec k} = \sqrt{\vec k^2+m^2}$\, and $V$ is 
the quantization volume. By performing the sum over $n$ 
in Eqs.(\ref{diagemt}) and (\ref{diagemt1}), we get the matter and the 
vacuum contributions to the  energy density 
$\rho= <<\hat T_{_{0\,0}}>>$ and pressure 
$p= <<\hat T_{_{i\,i}}>>$ (due to rotational invariance, 
$<<\hat T_{_{11}}>>=<<\hat T_{_{22}}>>=<<\hat T_{_{33}}>>$):

\bea\label{diagemt2}
\rho &=& \frac{1}{V}\sum_{\vec k} n_{_{BE}} \,\,
\omega_{\vec k} \,+\, \frac{1}{V} \sum_{\vec k}
\frac{\omega_{\vec k}}{2}
\equiv \rho^{\,m}  + \rho^{\,v} \\
p &=& \frac{1}{3V}\sum_{\vec k}\, n_{_{BE}} \,\,
\frac{{\vec k}^2}{\omega_{\vec k}}\,+\, \frac{1}{3V}\sum_{\vec k}
\frac{{\vec k}^2}{2\,\omega_{\vec k}}
\equiv p^{\,m}  + p^{\,v}\, , 
\label{diagemt3}\eea
where $n_{_{BE}}=n_{_{BE}}(\vec k^2,T)$ is the Bose-Einstein distribution
at temperature $T$. Again, the superscripts ``$m$'' and ``$v$'' 
are for ``$matter$'' and ``$vacuum$'' respectively.

The first terms in the r.h.s. of Eqs.(\ref{diagemt2}) and 
(\ref{diagemt3}), $\rho^{\,m}$ and $p^{\,m}$, 
come from the thermal average of the number operators\,
$a_{\vec k}^{\dagger} a_{\vec k}$\, and are the matter  
contribution to $T_{\mu\nu}$. It is worth to note that 
this is the only contribution usually considered in 
Eq.(\ref{einstein}): the energy momentum tensor of the 
relativistic gas of particles. On the other hand, 
$\rho^{\,v}$ and $p^{\,v}$ come from the thermal average of the
commutators $[a_{\vec k}^{\dagger},\, a_{\vec k}]$, i.e. from 
c-numbers, and coincide with the vacuum expectation values of 
the components of
$\hat T_{\mu\nu}$. Note also that $\rho^{\,v}$ is nothing but 
the term which is usually recognised as the zero point energy 
contribution to the cosmological constant. 
Eqs.(\ref{diagemt2}) and  (\ref{diagemt3}) provide an explicit 
example of the general relation shown in Eq.(\ref{enmomt}).

This elementary computation shows that 
the matter and the vacuum contributions to $T_{\mu\,\nu}$ 
do not come as separate entities. They are the result of 
a unique operation, namely the thermal average of the operator 
$\hat T_{\mu\nu}$ with respect to the Bose-Einstein distribution.
Both $\rho $ and $p $ contain on the same footing contributions from 
the matter and from the vacuum content of the theory. However, while 
the first terms in the r.h.s. 
of Eqs.(\ref{diagemt2}) and (\ref{diagemt3}) are convergent (due to 
the cutoff role played by the Bose-Einstein distribution), the second 
ones, i.e. the vacuum contributions, diverge. By explicitly performing 
the computation with the help of an ultraviolet cutoff we get:
\bea
\rho^{\,v}    =\frac{1}{16\pi^2}
\left[\Lambda(\Lambda^2+m^2)^{\frac32}-\frac{\Lambda m^2
(\Lambda^2+m^2)^{\frac12}}{2} 
-\frac{m^4}{4}  \ln\left(\frac{(\Lambda+
(\Lambda^2+m^2)^{\frac12})^2}{m^2}\right)\right]\,,
\label{rho}
\eea
\bea   p^{\,v}  =\frac{1}{16\pi^2}\left[
\frac{\Lambda^3(\Lambda^2+m^2)^{\frac12}}{3}
-\frac{\Lambda m^2
(\Lambda^2+m^2)^{\frac12}}{2} 
+\frac{m^4}{4}  \ln\left(\frac{(\Lambda+
(\Lambda^2+m^2)^{\frac12})^2}{m^2}\right)\right]\,.
\label{press}
\eea

By assuming that $\Lambda$ coincides with the Planck scale 
$\Lambda=M_{_P} >> m$, the ratio between $p^{\,v} $ and $\rho^{\,v} $ is 
essentially $1/3$: 
\be\label{mainv}
p^{\,v}  
\sim  \frac{\,\,\rho^{\,v}}{3} \,.
\ee
Moreover, when the matter content is relativistic, 
this is also the ratio between $p^{\,m} $ and $\rho^{\,m} $ and 
the EOS for the field $\phi$ is:

\be\label{main}
p = p^{\,v} +  p^{\,m} 
\sim  \frac{\rho^{\,v} + \rho^{\,m}}{3} = \frac{\rho}{3} \,.
\ee

These results are totally different from the usual ones, where for 
the vacuum component one has $p^{v} = - \,\rho^{v} $, i.e. a value 
of $w$ which is different from the matter one. 
As we have already noted, if we manage to get rid of the quartic 
and quadratic divergences with the help of some formal 
regularization procedure, the remaining terms in $p^{\,v}$ and 
$\rho^{\,v}$ would obey the usual vacuum equation  of state with 
$w=-1$. We also note that, as in Eq.(\ref{main}) $w$ turns out to 
be $\sim 1/3$, the above finding does not change the well known scaling 
of $\rho^{m}$. 

So far we have considered the simple example of a free theory (see 
Eq.(\ref{phiemt})).
However, these same steps can be repeated for any, even interacting,
field theory. Of course the presence of interaction terms such 
as $g\phi^4$ induces corrections to the Lagrangian parameters. 
In the case of mass, for instance, these corrections are 
proportional to $g\Lambda^2 + O(g^2)$. As long as $g$ is 
perturbative, we expect  these terms not to spoil the above 
analysis.

\section{Time evolution of the vacuum energy density}

Let us now consider the consequences of the above findings
for the cosmological constant problem. Being $w \sim 1/3$\,,\,
$\rho =\rho^{\,v} +\, \rho^{\,m}$ has the well known time evolution of 
the relativistic matter which is governed by the continuity and 
Friedman equations:

\bea\label{fried1}
&&{\dot \rho}  + 3\,\left(\frac{\dot a}{a}\right)
\left(\rho  +  p \right)= 0  \\
&&\left(\frac{\dot a}{a}\right)^2 = \frac{8 \pi G}{3} \rho\label{fried2}\,,
\eea
where $a(t)$ is the cosmic scale factor 
(consistently with the present observations, we have
considered a flat space, $k=0$). Note also that in Eq.(\ref{fried2}) 
we have neglected the ``classical'' (i.e. not originated 
from quantum vacuum fluctuations) $\lambda$ term in 
the Einstein equation (\ref{einstein}). As is well known,
the solution of Eq.(\ref{fried1}) is:
\be\label{sol}
\rho (t) \propto a(t)^{-4}\,.
\ee

Although Eq.(\ref{fried1}) and the corresponding solution (\ref{sol}) 
are obtained for $\rho=\rho^{\,v} + \rho^{\,m} $, we expect them to 
hold also for $\rho^{\,v} $ and $\rho^{\,m}$ separately. In fact, when 
no matter is present, $\rho$ reduces to $\rho^{\,v} $ so that 
Eq.(\ref{fried1}) is valid for $\rho^{\,v} $ alone. Then, if no substantial 
change in the behaviour of $\rho^{\,v} $ is induced by the presence of 
matter, $\rho^{\,m} $ satisfies Eq.(\ref{fried1}) too. Such a time 
evolution of $\rho^{m} $ is nothing 
but the well known evolution of relativistic matter: in the 
usual treatment, it is obtained by neglecting $\rho^{v} $ in 
the continuity equation (\ref{fried1}).

At early cosmological times (and therefore at high temperatures $T$)
one has $T >> m$ (we have taken the Boltzmann constant $k_B=1$) and this 
corresponds to the radiation, i.e. relativistic matter, dominated era:
\be\label{cmb}
\rho^{m}  (t) = \frac{\pi^2}{30}\,T^4\, \propto a^{-4}\,. 
\ee
As we noticed above, as long as matter is relativistic, 
$\rho^{\,m} $ and $\rho^{\,v} $ have the same scaling 
($\rho^{\,m\,,\,v}\propto a^{-4}$) so that we can write 
\be\label{rat}
\rho^{\,v}(t)=\frac{\rho^{\,v}(t_P)}{\rho^{\,m}(t_P)}\, \rho^{\,m}(t)\, , 
\ee
where we have chosen as initial time\, 
$t=t_P$\,, with $t_P=(M_P)^{-1}$, the Planck time.
Moreover, from Eq.(\ref{cmb}) we have that $\dot a/ a = - \dot T/ T$ and
Eq.(\ref{fried2}) can be written as:
\be\label{fried}
\left(\frac{\dot T}{T}\right)^2 = \frac{8 \pi G}{3}
\left(1+\frac{\rho^{\,v}(t_P)}
{\rho^{\,m}(t_P)} \right) \rho^{\,m} (t) =
\frac{4 \pi^3 G}{45}
\left(1+\frac{\rho^{\,v}(t_P)}
{\rho^{\,m}(t_P)} \right)\,T^4\,,
\ee
By integrating the above equation we get:
\be\label{tvst}
T=\left(\frac{45}{16\pi^3\,K\,G}\right)^{\frac14}\,t^{-\frac12}\,,
\ee
with  $K=1+{\rho^{\,v}(t_P)}/{\rho^{\,m}(t_P)}$. 
Note that in the standard approach, where
$\rho^{\,v}  (t)$ is not taken into account in the Friedman
equation (\ref{fried2}),  $K=1$.

Let us consider now the theory defined at the Planck time, 
$t_P$. If the cutoff is taken to be 
at the Planck scale, $\Lambda=M_P=1.22 \times 10^{19} GeV$, the leading
contribution to the vacuum energy density at $t_P$ is:
\be\label{unwanted}
\rho^{\,v} (t_P)=\frac{M_P^4}{16 \pi^2}\,. 
\ee 
From Eqs.(\ref{cmb}) and (\ref{tvst}) we then find:
\be\label{ratio2}
\frac{\rho^{\,m}(t_P)}{\rho^{\,v}(t_P)}=
\frac{3\pi}{2} - 1 \,\sim \, 3.71\,,
\ee
where we have used $G=M_P^{-2}=t_P^2$. In passing, we note that 
from Eq.(\ref{ratio2}) we have that $K\sim 1.27$. When this value 
of $K$ is inserted in Eq.(\ref{tvst}), we get a slight correction 
to the result obtained in the standard approach, where $K=1$.

The relevance of the result contained in Eq.(\ref{ratio2}), however, 
lies elsewhere. In fact, Eq.(\ref{rat}) predicts that, as long as matter 
is relativistic, the ratio $\rho^{\,m}(t)/\rho^{\,v}(t)$ is constant and  
given by Eq.(\ref{ratio2}). 
In particular, if we consider a massless field which is relativistic 
at any time, this ratio keeps such a value up to the present time 
$t_0$. Therefore, $\rho^{\,m}(t_0)$ is about four times $\rho^{\,v}(t_0)$. 
As the background photon density $\rho_\gamma(t)$ follows
precisely this scaling,
we find that: 
$\rho_\gamma(t_0) \sim 4 \, \rho^{\,v}(t_0)$. Therefore, since we know
that at present time\, $t=t_0$\, the contribution of 
$\rho_\gamma(t_0)$ to the total energy density is negligible,
the same must hold true for $\rho^{\,v}(t_0)$.
 
As $T$ decreases, matter evolves towards the non-relativistic 
regime (opposite limit, $T << m$) where \, $\rho^{\,m} \propto a^{-3}(t)$, 
while $\rho^{\,v}$ continues to follow its previous  scaling, 
$\rho^{\,v} \propto a^{-4}(t)$. 
During this epoch, the expansion of the universe, i.e. its scale 
factor $a(t)$, is controlled by non-relativistic matter so that, 
starting from $t=t_{eq}$, when
$\rho_{rel}(t_{eq})=\rho_{nrel}(t_{eq})$, the scaling of $\rho^{\,v}$ 
with $t$ changes.  

It is not difficult to estimate the value of $\rho^{\,v}$ at the present 
time $t_0$. The computation goes as follows. By integrating Eq.(\ref{fried1}) 
for $\rho^{\,v}$  from $t_P$ down to $t_{eq}$, i.e. during the radiation era, 
as $a(t) \sim \, t^{1/2}$ we get:  
\be\label{rhov1}
\rho^{\,v}(t_{eq}) = \rho^{\,v}(t_P)\left(\frac{t_P}{t_{eq}}\right)^2\,. 
\ee
During the successive period, the matter
dominated era, it is still $\rho^{\,v} \propto a^{-4}$, 
but now $a(t)\sim \, t^{2/3}$. Therefore, by integrating Eq.(\ref{fried1}) 
for $\rho^{\,v}$ from $t_{eq}$ down to $t_0$ we have: 
\be\label{rhov2}
\rho^{\,v}(t_0) = \rho^{\,v}(t_{eq})\left(\frac{t_{eq}}{t_0}\right)^{\frac{8}{3}}\,,
\ee
so that, at the present time, $\rho^{\,v}(t_0)$ is: 
\be\label{rhov4}
\rho^{\,v}(t_0) = \rho^{\,v}(t_P)\left(\frac{t_P}{t_0}\right)^2\cdot
\left(\frac{t_{eq}}{t_0}\right)^{\frac23} =
\rho^{\,v}(t_P)\left(\frac{t_P}{t_0}\right)^2 \cdot \frac{a_{eq}}{a_0}\,\, .
\ee
By inserting now in Eq.(\ref{rhov4})\,  $\rho^{\,v}(t_P)$ given in 
Eq.(\ref{unwanted}), $t_P\sim 5 \times 10^{-44}\,s$, $t_0 \sim 2/(3 H_0)$, with 
$(H_0)^{-1} \sim 13.7$ Gy and ${a_{eq}}/{a_0} \sim 1/3048$ 
\cite{pdg}, we finally  find: 
\be\label{rhov3}
\rho^{\,v}(t_0) \sim \left ( 1.93 \times 10^{-4} \, {\rm eV}  
\right )^4\, .
\ee
We would like to compare now this result for $\rho^{\,v}(t_0)$ with 
the determination of $\rho_{\gamma}$ at present time\cite{pdg},
\be\label{rhogam}
\rho_{\gamma}(t_0) \sim \left ( 2.11 \times 10^{-4} \, {\rm eV}  
\right )^4\, .
\ee

As can be easily checked, compatibly with the 
numerical uncertainties of the various  quantities involved,
the ratio between $\rho_{\gamma}(t_0)$ 
and $\rho^{v}(t_0)$ is in substantial agreement with the 
prediction of Eq.(\ref{ratio2}).
As photons are always relativistic, this is precisely 
what should be expected from our previous analysis.  
In fact, as the measure of $\rho_{\gamma}$ is an experimental input 
totally independent from our analysis, we can consider this finding 
as a check on our ideas. Moreover, Eq.(\ref{rhov3}) shows that, 
as is the case for photons, the contribution of $\rho^{\,v}$ is 
nowadays negligible.
 
To summarize, we suggest that the cosmological evolution itself 
provides the mechanism that dilutes the zero point 
energy contribution to the total energy density of the universe
down to a value which is negligible if compared to the current 
matter and cosmological constant determinations. 

Another interesting outcome of our analysis is the following. As 
already noted, when the energy momentum tensor of the vacuum is 
not of the form $T_{ \mu\nu}\propto g_{ \mu\nu}$, the Lorentz 
invariance of the theory is lacking. Our $<\hat T_{ \mu\nu}>$ 
at Planck time has not a Lorentz invariant form, but the cosmic 
evolution allows to recover Lorentz invariance at our time. 
We think that the connection between our findings and the whole 
subject of Lorentz violation at Plank scale is worth of further 
investigations.

Before ending this section, it is worth to spend some additional 
words on the underlying field theoretical setup of our work. As 
is clear from the previous section, up to now we have considered 
a Fock space in a flat Minkowski space-time. Clearly, a more 
rigorous treatment of the problem would have required the use of  
quantum field theory (QFT) in an expanding universe, as is the  
case (of interest for us) of a Friedmann-Robertson-Walker (FRW) 
background. 

As we shall show in a moment, however, it is not difficult to 
convince ourselves that such a refinement is irrelevant for the 
issue under investigation. Actually, we have deliberately chosen 
to work on a flat space-time since our goal is to present the 
mechanism of the washing out of the zero point energies of the
effective field in the simplest possible framework, avoiding any 
unnecessary technical detail. 

In fact, let us consider a scalar quantum field in a FRW 
background, a problem largely investigated in the literature 
\cite{park1, park2, ful, birrdav}. 
Regularization procedures based on ``point splitting'' or on
``adiabatic regularization''  both  give the same result for 
the leading divergences in the vacuum pressure and density, namely  
$p^v=\rho^v/3$.  

Clearly, from our effective field theory point of view, the 
``adiabatic basis'' approach \cite{park1, park2}, which allows 
for a mode decomposition, is the most appropriate. In fact, 
this property allows for the definition of a Fock space at 
each time, similarly to what happens in the flat case. Moreover,
it is easy to see that the leading ``divergent'' term of the 
vacuum energy density scales as  $\rho^v \sim a(t)^{-4} \Lambda^4$, 
where $a(t)$ is the scale factor in the FRW metric and $\Lambda$ 
is the UV cut-off. This is nothing but our result.

In this respect, it is important to stress once again that 
our results are derived in the framework of an Effective Field 
Theory approach. This is completely different from a renormalized 
theory, which is the point of view considered in the above mentioned 
literature, where the divergent terms are treated as unphysical 
and are, therefore, cancelled out. 
In our Effective Theory approach the physical cut-off is part of the 
definition of the theory itself and plays an important role in 
establishing the physical results. In such a framework, 
the cut-off dependence of $\rho^v$ and $p^v$ is an essential 
physical aspect of our analysis.

\section{The counting of the degrees of freedom}

Up to now we have considered the cosmological evolution of the
(thermal average of the) energy-momentum tensor of a quantum
scalar field starting at the Planck time $t_P$,
with  the  assumption that at $t \sim t_P$ and
$E\sim M_P$ physics is entirely described by one
quantum field (or a small number of fields) and that the
known lower energy theories were born during the cosmic time
evolution\footnote{As we have already said, a different
(probably string) theory is supposed to describe the physics at
times earlier than $t_P$.}.
This assumption appears natural in view of our ideas on
the effective nature of particle physics theories and fits our 
current views on the cosmological evolution. In this respect, 
the lower energy new fields, new degrees of freedom (dof), are 
nothing but
a convenient manner to parametrise the theory at a lower scale.
Therefore, when computing the vacuum contribution to the
cosmological constant, one should not include the zero point
energies of the effective low energy theories as this would
result in a multiple counting of dof. The zero
point energies coming from the dof of the original
quantum field already account for the whole contribution to the
vacuum  energy.

Before we can conclude that our findings can be of some relevance for the
cosmological constant problem, we still have to address another
issue. As is well known, some of our low energy theories,
for instance the Higgs sector of the Standard Model, are
characterised by the presence of condensates. In the standard approach,
these terms are considered to give very large contributions to the
cosmological constant as 
they  enter the energy momentum tensor as 
$\rho_{c}\,g_{\mu\nu}$, where $\rho_{c}$ is the vacuum energy 
density associated with the condensate.
However, according to our previous discussion,
there are no such additional terms as the whole
contribution is already contained in the zero point energies of the
original theory. Again, taking into account these terms would result
in a double counting of dof.
A similar point of view has already been expressed 
within a different approach  to the cosmological constant problem \cite{volovik}. 

Below we try to elucidate the arguments 
of the previous two paragraphs with the help of an example
inspired to the work on the top quark condensates of
Bardeen, Hill and Lindner\cite{bardeen}.

Following \cite{bardeen}, let us consider a Nambu
Jona-Lasinio theory defined at the high energy scale $\Lambda$ by:
\be\label{njl}
Z=\int\,D\bar\psi \, D\psi\,
{\rm  
exp}\,\left[\,i\int\,d^4\,x\,\left(\bar\psi(i\gamma^\mu\partial_\mu -  
M)\psi +
\frac{g^2}{2 m^2_0}\bar\psi\psi\bar\psi\psi \right) \right]\,.
\ee
An Hubbard-Stratonovic transformation introduces a new scalar field
$\phi$ so that Eq.(\ref{njl}) can be rewritten as:
\be\label{njlh}
Z=\frac{1}{\cal N}\int\,D\bar\psi \, D\psi\, D\,\phi\,
{\rm exp}\,\left[\,  
i\int\,d^4\,x\,\left(\bar\psi(i\gamma^\mu\partial_\mu - M)\psi
-\frac{m^2_0}{2}\phi^2 + g\bar\psi\psi\phi\right) \right]\,,
\ee
where the normalisation factor ${\cal N}$ ensures the equality of
Eqs.(\ref{njl}) and (\ref{njlh}). Obviously, any quartic divergent term
which apparently comes from the zero point energies of $\phi$ cannot induce
any change in the quartic divergences of Eq.(\ref{njl}) as they are
cancelled by ${\cal N}$.

The next step in \cite{bardeen} consists in the integration of the
high frequency modes of the fermion and scalar fields from 
$\Lambda$ to the lower energy scale $\mu$:
\bea\label{njlhl}
Z=\frac{\cal Q}{\cal N}&\int&D\bar\psi_l \, D\psi_l \, D\,\phi_l\,
{\rm exp}\,\Bigl[\, i\int\,d^4\,x\,
\Bigl( \bar\psi_l(i\gamma^\mu\partial_\mu - M - \delta M)\psi_l 
\nonumber \\
&&+ g\bar\psi_l\psi_l\phi_l  +\frac12  
Z_\phi\partial^\mu\phi_l\partial_\mu\phi_l
-\frac{m^2_0 +\delta m_0^2}{2}\phi_l^2
- \frac{\lambda}{24}\phi_l^4\Bigr) \Bigr]
\eea
where $\phi_l$ and $\psi_l$ are the scalar and fermion fields
with Fourier components up to $\mu$. This integration generates  
new dynamical degrees of freedom \cite{eguci}
in the Lagrangian of Eq.(\ref{njlhl}).

This example is relevant to our problem for the following reason. 
When one deals with the effective Lagrangian of Eq.(\ref{njlhl}),  
the normalisation factor ${\cal Q}/{\cal N}$ is not considered
as one has no knowledge of the higher energy theory. Clearly, this 
has no effect in the evaluation of the low energy Green's functions, 
i.e. for typical scattering processes. However, if we compute the 
vacuum energy from the quartic divergences of this effective 
Lagrangian, we end up with a result which differs from the one 
obtained from the ``fundamental'' theory of Eq.(\ref{njl}) because of an
erroneous counting of the dof. Only if we take into account the 
normalisation factor ${\cal Q}/{\cal N}$ we recover the 
original result. Clearly, the same argument applies when additional 
contributions to the vacuum energy come from the appearance of 
condensates such as, for instance, a vacuum expectation value for 
$\phi_l$. 

We can also consider an alternative, but equivalent, argument which
allows to understand the suppression of the $\Lambda^4$ and the  
condensate terms. 
Let us consider the appearance of a condensate below some 
temperature $T_{SB}$ through a  symmetry breaking mechanism. 
The cutoff of the  low energy theory which 
describes the broken symmetry phase is nothing but the temperature 
$T_{SB}$ at which the transition takes place. Moreover, the cutoff and the condensate
contributions to $\rho^{\,v}$ and $p^{\,v}$ come in the same combination 
as in Eqs.(\ref{rho}) and (\ref{press}), where the $m^4$ terms are 
now accompanied by the additional $v^4$ condensate contribution
($v$ is the value of the condensate). 
As is always the $\Lambda^4=T^4_{SB}$ term which dominates, we obtain for 
$\rho^{\,v}$ the same scaling as before, regardless of the 
Lorentz invariant nature of the condensate contribution. 
Being $T_{SB}$ the cutoff, again we find that these contributions at 
present time are suppressed. 

\section{Summary and conclusions}

We have found that if we consider that at the 
Planck time $t_P$ physics is described by an effective field 
theory with ultraviolet cutoff $M_P$, the corresponding vacuum 
energy density undergoes a cosmic scaling that makes it 
negligible at present time $t_0$ when compared to non-relativistic 
matter and cosmological constant densities, much in the same way 
as the cosmological scaling makes the photon density negligible 
nowadays. The reason for this behaviour is that for an effective 
field theory $<\hat T_{\mu\nu}>$ is such that $p^{v} \sim \rho^{v}/3$. 

Moreover, our analysis predicts a constant ratio, Eq.(\ref{ratio2}), 
between the vacuum and the radiation densities. When 
the theoretical determination of the vacuum energy density at  
present time, given in Eq.(\ref{rhov3}) and obtained by a proper rescaling 
of the Planck time vacuum density of Eq.(\ref{unwanted}), is compared with 
the experimentally determined photon energy density in Eq.(\ref{rhogam}), 
we find substantial agreement with our prediction. 

We believe that this supports the central idea put forward in the present  
work, namely that zero point energy and condensate contributions to the 
universe energy density are washed out by the cosmological evolution.
Moreover, these terms, being $w \sim 1/3$, cannot contribute to the 
cosmological constant, for which we know that the measured value of $w$
is $w\sim -1$. In our opinion, this result points towards a gravitational 
origin of the (measured) cosmological constant.

\end{document}